\documentclass[12pt]{JHEP3}
\usepackage{graphicx,amsmath,amssymb}
\usepackage{epsfig,multicol}
\usepackage{epsf}
\usepackage{epstopdf}
\input{epsf.sty}

\def\m10{M_{10}}

\def\baray{\begin{eqnarray}}
\def\earay{\end{eqnarray}}

\newcommand{\be}{\begin{equation}}
\newcommand{\ee}{\end{equation}}
\newcommand{\bea}{\begin{eqnarray}}
\newcommand{\eea}{\end{eqnarray}}
\newcommand{\barr}{\begin{array}}
\newcommand{\earr}{\end{array}}

\newcommand{\ba}{\begin{eqnarray}}
\newcommand{\ea}{\end{eqnarray}}
\newcommand{\D}{\overline{\mbox{D}}}

\newcommand{\simless}[0]{\mathbin{\lower 3pt\hbox
   {$\rlap{\raise 5pt\hbox{$\char'074$}}\mathchar"7218$}}}
\newcommand{\simgreat}[0]{\mathbin{\lower 3pt\hbox
   {$\rlap{\raise 5pt\hbox{$\char'076$}}\mathchar"7218$}}}
\newcommand\msun{{\rm M}_\odot}

\title{Cosmic String Detection via Microlensing of Stars}
\author{David F. Chernoff $^{1}$ and S.-H. Henry Tye $^2$
\\ \small{\em $^1$Department of Astronomy \\
Cornell University, Ithaca, NY 14853} 
\vskip .1cm
\\ \small{\em $^2$Newman Laboratory for Elementary Particle Physics \\
Cornell University, Ithaca, NY 14853} }

\abstract{ Cosmic superstrings are produced towards the end of the
brane inflation, a scenario realized in modern superstring theory.
If the string tension is low enough, loops tend to be relatively long-lived.
The resultant string network is expected to contain many loops which
are smaller than typical Galactic scales. Cosmic expansion damps the
center of mass motion of the loops which then cluster like cold dark
matter, eventually decaying by emission of gravitational radiation.

Loops will lens stars within the galaxy and local group. We explore
microlensing of stars as a tool to detect and to characterize some of
the fundamental string and string network properties, including the
dimensionless string tension $G \mu/c^2$ and the density of string
loops within the Galaxy.

As $G \mu \to 0$ the intrinsic microlensing rate diverges as $1/\sqrt{G
\mu}$ but experimental detection will be limited by shortness of the lensing
timescale and/or smallness of the bending angle which each vary
$\propto G \mu$. We find that detection is feasible for a
range of tensions. As an illustration, the planned optical astrometric
survey mission, Gaia, should be able to detect numerous microlensing
events for string networks with tensions $10^{-10} \simless G \mu
\simless 10^{-6}$. A null result for optical microlensing implies $G
\mu \simless 10^{-10}$.

If lensing of a given source {\it is} observed it will repeat because the
internal motions of the loop are relativistic but the center of
mass motion may be much smaller, of order the halo velocity $v_h$. 
This distinctive hallmark, $c/v_h \sim 1000$ repetitions, 
suggests a useful method for confirmation of a potential lensing
detection.

We argue that the estimate of the Galactic lensing frequency is likely
to rise with more realistic descriptions of the superstring network
while the effect of the inclusion of loop-loop interactions within
the Galactic halo is not yet clear.

 }
\preprint{astro-ph/yymmnnn}

 
\begin{document}

\section{Introduction}

Inflation probably explains the origin of our universe \cite{guth}.  A
concrete realization of this scenario in superstring theory is brane
inflation \cite{Dvali:1998pa,collection,Kachru:2003sx}.  The simplest
version involves an interacting D3-${\D}$3-brane system in which the
D3-brane moves towards the ${\D}$3-brane sitting at the bottom of a
warped throat, with the position of the mobile D3-brane corresponding
to the inflaton.  Inflation ends when the D3-brane and the
${\D}$3-brane collide and annihilate, initiating the hot big bang.
Cosmic strings are copiously produced during the epoch of annihilation
\cite{Jones:2002cv,Sarangi:2002yt}. These strings are simply
superstrings (F- and D-strings and their bound states
\cite{Dvali:2003zj,Copeland:2003bj,Firouzjahi:2006vp}) stretched to
cosmological sizes.  Finding evidence for the existence of these
strings would go a long way towards confirming certain fundamental
aspects of superstring theory \cite{Tye:2006uv,Polchinski:2007qc}.

The single most important property of a cosmic string is its tension
$\mu$, or, in dimensionless terms, the characteristic gravitational
potential $G \mu/c^2$ \cite{Vilenkin} (hereafter, we will take $c=1$). An
initial estimate of the range of tension produced towards the end of
inflation gave $10^{-11} \simless G \mu \simless 10^{-6}$
\cite{Sarangi:2002yt}.  A recent analysis of the reheating that accompanies
brane inflation, including multi-brane multi-throat scenarios,
suggests that $G \mu < 10^{-11}$ is quite possible \cite{Chen:2006ni}.
The properties of the cosmic superstrings produced at the end of
inflation are compatible with all observations today but are beginning
to be constrained by observational data
\cite{smoot,Jenet:2006sv,Siemens:2006yp}.  The actual bound
is sensitive to the details of the string tension spectrum and the
probability of string-string interactions but $G \mu \simless 10^{-7}$
gives a rough idea of current observational constraints.

For small $G \mu$, it will be very challenging to detect cosmic string
signatures in lensing, gravitational wave bursts, pulsar timing and B
mode polarization in the cosmic microwave background radiation by
methods that have been investigated hitherto.  However, even low
$G\mu$ strings produce double images of point-like objects. Here, we
will investigate the utility of microlensing of stars to learn about
the string content of the universe.  The propensity to microlens
depends, of course, upon the number of lenses and their cross section
for bending light. Our current understanding of string network
evolution implies that the rate of lensing increases as $G\mu$
decreases. This happy circumstance may eventually lead to practical
experiments to detect superstrings. Microlensing of stars is much more
promising than microlensing of quasars \cite{Kuijken:2007ma}.

Although it is well accepted that cosmic strings evolve to a scaling
network \cite{Albrecht:1984xv}, string network evolution is by no
means well understood.  Even if one had a complete knowledge of the
intrinsic string tensions in string theory, the cosmological evolution
of the string network remains a challenging problem
\cite{Polchinski:2007qc}. Recent analyses strongly suggest that cosmic
superstrings evolve dynamically so as to produce a scaling solution in
which there exists a stable relative distribution of strings with
different quantum numbers \cite{Jackson:2004zg,Tye:2005fn}.  This is
very much like the behavior of cosmic strings generated by abelian
Higgs or Nambu-Goto type models \cite{Albrecht:1984xv,Vilenkin}.  The
superstring solutions found in current string theory, however, form a
much larger class which is only now being explored.

A scaling solution implies the fractions of the critical energy density
in long strings and loops are constant.  In fact, the determination of
$\Omega_{long}$ and $\Omega_{loop}$ has been an
outstanding question for some time \cite{Vilenkin}. 
Numerical simulations for cosmic strings may be fitted by
\be
\label{fdensity}
\Omega_{s} \equiv \Omega_{long} + \Omega_{loop} = \psi G \mu + \chi \sqrt{G \mu}
\ee
for constants $\psi$ and $\chi$. A small isolated loop of length $l$
will decay in a characteristic time $\tau = l/(\Gamma G \mu)$ and
$\Gamma \sim 50$ for Nambu-Goto strings. A common assumption made in
many simulations is that small loops are produced by the intersection
of long strings (including self intersection). For $H \tau << 1$ where
$H$ is the Hubble constant the loops decay quickly via gravitational
radiation and $\Omega_s \sim \Omega_{long} \sim \psi G \mu$; the
loop contribution is negligible. Numerical simulations show $\psi \sim
\Gamma$. Presumably, this is a good approximation for $G \mu$ close to
the present day observational bound.

Several new considerations have emerged from the latest simulations and
from theoretical work.
\begin{itemize}

\item{Loops dominate:} For small $G \mu$ newly
formed loops do not immediately dissipate. The latest simulations 
\cite{Olum:2006at, Ringeval:2005,Martins:2005} imply a loop
distribution which, if truncated by gravitational wave damping, yields
a qualitatively new effect: $\Omega_s \sim \Omega_{loop} >> \Omega_{long}$.

\item{Small scales on long strings:} Analytic work
\cite{Polchinski:2006ee,Polchinski:2007qc} shows that the stretching
of long strings is insufficient to prevent the development of small
scale structure. Collisions of long strings may produce large numbers
of loops of all sizes. Estimates indicate $\sim 10$\% of the long string
length ends up in loops with
scale comparable to the horizon. Loops with $H \tau \simgreat 1$ contribute
to $\Omega_s$.

\item{Backreaction of small scales:} When $\Omega_{loop}$ increases,
small loops can recombine to form bigger loops and, likewise, isolated
loops may be incorporated into long strings.  The determination of the
scaling solution when energy can flow up and
down a hierarchy of scales is an outstanding problem.

\item{Damping of loops vs long strings:} Analytic work
\cite{Siemens:2001,Siemens:2002} indicates that the rate of damping of
modes on long strings differs from $\Gamma G \mu$, the characteristic
loss rate for an isolated loop. Under such circumstances the network-
and time-averaged gravitational energy loss rate per mode is coupled
to the details of the intercommutation process.

\end{itemize}

These factors motivate the development of a very approximate
description of the string network when the tension is
low. Specifically, we extend the analysis of Olum et al. in
Ref.\cite{Olum:2006at} to describe the loop population for small $G
\mu$. We also parameterize the effective decay rate of loops in the
network by $\Gamma_{R} G \mu$ where, in general, we expect $\Gamma_{R}
< \Gamma \sim 50$.

This minimal quantitative framework is used to begin an investigation
of microlensing by loops. Lensing by cosmic string loops at
cosmological distances was considered originally in
\cite{Vilenkin84,HoganNarayan84}. The analyses focused on
$G \mu \sim 10^{-6}$ and the possibility that the lensing might
produce multiple images of QSOs. The situation of interest here
is quite different though the physical process, gravitational lensing,
remains the same as was originally discussed in these
papers. We shall first give a brief sketch of the
latest status of the cosmic string network. We shall outline the form
of the loop distribution which is presented in full detail in the
Appendix.  We shall then use the result to estimate the event rates
and the detectability of microlensing for the cosmic strings. While
many properties of superstrings and astrophysics remain unaccounted
for (specifically, the intercommutation probability and the
collisional interactions within the Galaxy) this approach highlights
the role of lowering $G \mu$ and sets a benchmark that we will refine
as the input physics is better understood.

Our analysis shows that the intrinsic rate of microlensing increases
as tension decreases. At the same time the detection efficiency for a
microlensing event diminishes because both the duration of the event
and also the size of the deficit angle diminish.\footnote{ The deficit
angle due to the string must be bigger than the angle subtended by a
star.  For typical sources (a solar mass main sequence star within the
halo) this is roughly $G \mu \simgreat 10^{-14}$.  Likewise, the
duration $\delta t_{lens}$ of the shortest detectable optical
microlensing event is limited by, among other things, the ability to
detect a factor of 2 change in the number of the source photons.  If
we assume a typical source (a solar mass star at distance $10$ kpc),
a broadband instrument detecting photons near the peak of the spectrum
($\sim 1$ eV), a typical instrument (a 10\% efficient meter-class
telescope), then we can estimate that a 5$\sigma$ detection requires
$\delta t_{lens} \simgreat 40$ ms. This corresponds to roughly $G \mu
\simgreat 10^{-14}$.}

To illustrate the potential utility for detecting cosmic string
microlensing of stars, we consider the upcoming European Gaia mission
\cite{Gaia}. Gaia is primarily an astrometry satellite designed to
measure proper motions and parallaxes of about $10^9$ stars in and
near the Galaxy over a 5 year period. The satellite records the flux
for each star and it is this feature which is of greatest interest to
us.  On average each star will be observed 80 times; individual
observations take about $\sim 3$ seconds.  The expected accuracy of
the photometry from a single observation is 
easily sufficient to see a factor of 2 change in flux from one
observation to the next.  Using a crude model of detection efficiency,
{\it we find that an instrument like Gaia would be able to detect
microlensing events for cosmic strings in the Galaxy: a few events for
$G\mu \sim 10^{-10}$ to a few dozens for $G\mu \sim 10^{-8}$.}  This
estimate invokes several conservative choices including
the value of $\Gamma_{R}$;
the event rate increases for $\Gamma_{R} < \Gamma$.

A key reason for this detectable rate is the over-density of loops in
the Galaxy. Even small loops can survive a Hubble time when the
tension is low. A generally new scenario for cosmic string evolution
now unfolds: the velocity of the center of mass of a loop decreases on
account of cosmic drag; loop-loop interactions freeze out.  Loops
subsequently behave like cold dark matter, slowly radiating
gravitational waves. Most of the loops of interest for microlensing
were born during the radiation era and fall into gravitational
perturbations that begin to grow after equipartition.  In particular,
the loops track the perturbations of the cold dark matter. This
clumping enhances the local loop density by at least a factor of $\sim
10^{5}$.

From the string theory point of view, the true lensing rate may be
{\it enhanced} above our estimate because the intercommutation
probability of superstrings may be as small as $P \sim 10^{-3}$
\cite{Jackson:2004zg}.  Also, cosmic superstrings come in a variety of
tensions and charges so that a number of species are present in the
network \cite{Tye:2005fn}.  These effects will tend to increase the
energy density throughout the universe in the superstring network in eq.
(\ref{fdensity}) roughly like $ \Omega_{s} \rightarrow {n}
\Omega_{s}/P$, where $n$ is the effective number of types, $n \sim 5$.
The spectrum of cosmic superstrings yields a discrete set of
tensions that can easily vary by an order of magnitude.

From an astrophysical point of view, the true lensing rate may be 
{\it enhanced} by dissipative loop-halo interactions that boost
the galactic loop density and it may be {\it
diminished} by the collisional interaction of loops once they begin to
clump in the Galaxy. Intercommutation might chop up
small loops into smaller loops and shorten their lifetime
to gravitational
wave emission. This, in turn, may diminish the part of the
spectrum responsible for most of the lensing while increasing the
locally generated gravitational wave background. We have not included
either the dissipation or the interaction 
in our microlensing estimates but will return to these issues
in the future.

\section{String Network}
 
We shall start by considering cosmological cosmic strings like
Nambu-Goto strings or vortices in an abelian Higgs model. One readily
identifies two components, long horizon-crossing segments and
sub-horizon closed loops.  Only straight strings are static; all
others are dynamic with relativistic motion. Even isolated loops
oscillate in a highly complicated fashion.  A network of long strings
and loops can change topology by the process of intercommutation, the
breaking and rejoining experienced when the motions of two segments of
string cause them to coincide in 3D space.  This process can fragment
as well as rejoin the basic elements of the string network. Self
intersections of a long string cuts off new loops, self intersections
of a loop transforms it to multiple loops, etc.  Conversely, inverse
processes allow loops to reconnect to long strings and loops to
reconnect with other loops. In the cosmological context the processes
of intercommutation, damping (e.g. gravitational wave emission) and
cosmological expansion govern the string network evolution.

When a network exhibits scaling behavior the energy density (either
long or loop) is a fixed fraction of the critical energy density eq.
(\ref{fdensity}). When very small loops are formed and decay
promptly the critical density
\be
\Omega_s \sim \Omega_{long} \sim  \psi G \mu
\ee
and simulations show $\psi \sim \Gamma$ where $\Gamma \sim 50$
for abelian strings.
The scaling solution depends upon $\Gamma$ in an indirect way. Small
scale structure on the long strings is damped by gravitational wave
emission with rate $\propto \Gamma G \mu$ if one assumes modes damp
like those of an isolated loop.  An intersection of two long string
segments is more dissipative if it converts a greater overall length
to small loops. In fact, more loops are produced if the segment
possesses more small scale structure. The dissipation per collision of
long string segments, therefore, scales $\propto 1/\Gamma G \mu$.  The
rate for one segment of a horizon-crossing string to encounter another
$\propto \Omega_{long}$.  Therefore, the total dissipation rate per
long string segment $\propto \Omega_{long}/\Gamma G \mu$.  A fixed
rate of dissipation per string is essential to achieve a scaling
solution in the first place, so that $\Omega_{long} \propto \Gamma G
\mu$.

The specific scenario above depends on the $H \tau << 1$
where $H$ is the Hubble constant, $\tau = l/\Gamma G
\mu$ is the damping time and $l$ is the loop size. 
In a scaling solution a typical loop formed
at time $t$ has size $l \sim \alpha t$ where $\alpha$ is constant so
$\alpha/\Gamma G \mu << 1$ is required. The value of $\alpha$ is
poorly known; $\alpha < 10^{-12}$ is sometimes invoked \cite{Vilenkin}
but recent numerical simulations
\cite{Vanchurin:2005yb,Vanchurin:2005pa} and analytic studies
\cite{Polchinski:2006ee} suggest that loops with a range of sizes
$10^{-4} \simless \alpha \simless 0.25$ are created. Lower tension
implies longer damping time so that the prompt loop decay which is
characteristic of the scenario above is now more difficult to achieve.

To build a model capable of describing the loop distribution for
small $G \mu$ we follow the line of reasoning of
Olum et al. \cite{Olum:2006at}. In the radiation or matter-dominated
eras the number of loops produced per unit loop length per unit
volume per unit time has the form
\be
\label{ndef}
\frac{dN(l,t)}{dl dt dV} = t^{-5} f(x) \qquad \text{with $x=l/t$}
\ee
for some function $f(x)$ for a scaling solution.\footnote{This form
does not apply during the recent, $\Lambda$-dominated phase since the horizon
is not $\propto t$. The loops of
direct interest to us are from earlier epochs.} Numerical simulation in
Ref.\cite{Vanchurin:2005yb,Vanchurin:2005pa} suggest
a power law distribution of loops
\be\label{eqn:f}
f(x) = A x^{-\beta} \qquad \text{for $x<\alpha$}
\ee
where $\alpha t$ is the largest loop scale and $\beta < 2$. The
constants $A$, $\alpha$ and $\beta$ may be extracted from time-dependent
simulations for cosmic strings which appear to have entered a scaling regime.

Assume that in each infinitesimal time interval $(t,t+dt)$ the network
produces the loops described with $l < \alpha t$ according to $dN/dl
dt dV$ and these are subsequently diluted by cosmological expansion
without further intercommutation. This is an instantaneous fragmentation
description adjusted to agree with the results of time-dependent simulations
that model the complete process. A non-interacting loop shrinks by
gravitational wave emission until it disappears in time $l/(\Gamma_R
G\mu)$. Note the introduction of $\Gamma_R$ in the dimensionless decay
rate in place of $\Gamma \sim 50$. Here, $\Gamma_R < \Gamma$ implies
that loops live longer because of complex unmodeled network
effects. For example, in the conventional view, a loop of size $l$
emits energy at half the total rate of two loops of size $l/2$.  In
addition, \cite{Siemens:2001,Siemens:2002} find that the damping rate
of long strings is $\propto (G \mu)^k$ with $k>1$. For the simple
energy loss recipes employed it makes a great deal of difference
whether a length is part of a small loop, a large loop or a long,
horizon-crossing string. We do not attempt to disentangle these
effects but simply introduce $\Gamma_R$ as an effective damping rate.

The superstrings differ from previously studied
cosmic strings in fundamental characteristics like $G
\mu$ and $P$ and possibly in the resultant loop distribution described by
$A$, $\alpha$ and $\beta$. It is tricky to extrapolate to describe
domains not previously simulated.  Our calculation of the distribution
function for loops is given in detail in Appendix 1. The general idea
is as follows: When the network scales the long strings are chopped to
loops whose total length is a constant fraction of the horizon. Energy
conservation implies that $\int f x dx$ is fixed. If the typical loop
size $\alpha$ were to be modified (for example, because the
intercommutation probability is decreased), then this energy
conservation argument gives $A \propto \alpha^{\beta-2}$.  Using this
reasoning we can, in principle, consider various descriptions for $f$,
cosmological expansion dynamics, $\Gamma_R$ and $G \mu$. The minimal
assumption is that $(1-\langle v^2 \rangle)/\gamma_s^2$ is the same
for undamped Nambu-Goto strings in Minkowski space as for damped
superstring networks in FRW cosmology; here, the characteristic
inter-string distance $d(t)=\gamma_s t$ and $\langle v^2 \rangle$ is
the square of the string velocity averaged over the length of long
strings.  For our applications, however, we also assume the
simulation-derived quantities like $\beta$ are fixed and
concentrate on changes to
the network that arise from varying $G \mu$, $\Gamma_R$ and the
cosmological dynamics.

At any time the distribution of non-interacting loops in a given
volume is the integrated production rate over the history of the
universe
\be
\frac{dN}{dldV} = \frac{1}{V} \int \frac{dN}{dldVdt'} V(t') dt'
\ee
subject to upper and lower cutoffs. The upper cutoff corresponds to
the time when gravitational wave damping removes the loop; the lower
cutoff is the earliest epoch when the loop might be produced.  If the
expansion scale factor varies $a \propto t^n$ then the integrand
$\propto t^{3n + \beta -5}$. For both radiation and matter dominated
eras, the loop distribution at length $l$ is typically dominated by production
at early times $t=l/\alpha$ (this requires $\beta < 7/2$ or $3$,
respectively) so that
\be
\label{eqn:dNdldV}
\frac{dN}{dldV} \propto \frac{l^{3n-4}}{t^{3n} \alpha^{3n-2}}
\ee
as previously demonstrated by Vanchurin et al. \cite{Vanchurin:2005pa}.
This distribution may then be compared to the cosmological simulation results
\cite{Ringeval:2005,Martins:2005}. None of the simulations incorporate
gravitational damping so that the cutoff at small loop size in the
simulations reflects either initial conditions, finite resolution
and/or finite simulation times. Nonetheless, the results
at various epochs illustrates the buildup of a powerlaw distribution
of loops at intermediate scales with the predicted slope. In the
radiation era, for example, eq. (\ref{eqn:dNdldV}) gives $dN/dldV \propto
l^{-2.5}$ while the simulations of Ringeval et al. \cite{Ringeval:2005} 
imply $dN/dldV \propto l^{-2.6}$.

The energy density, lensing probability and lensing rate of loops
all involve essentially the same moment of the loop distribution,
$\int l dl \frac{dN}{dldV}$. The cutoff $l = \Gamma_R G \mu t$
is the key parameter which varies with the tension when we evaluate
those quantities. For loops originally generated in the radiation era
\be
\rho_{loops} \propto 
\frac{\mu}{t^2} \left( \frac{\alpha}{\Gamma_R \mu} \right)^{1/2}
\ee
Since $\rho_{cr} \propto 1/Gt^2$ we infer
\be
\Omega_{loops} \propto \sqrt{\frac{\alpha \mu}{\Gamma_R}}
\ee
This illustrates the combination of parameters that determines
$\Omega_{loop}$; if $\Omega_{long} \propto G \mu$ this
square-root behavior suggests that
$\Omega_{loop} > \Omega_{long}$ for small $G \mu$ .

The numerical simulations which include both loops and long strings
allow a quantitative check. Assume that the slope and amplitude of
the simulation-derived loop distribution is extended to the
gravitational cutoff and that the long string density is fixed.
The ratio is
\be
\frac{\Omega_{long}}{\Omega_{loop}} = \left\{
\begin{array}{cc}
4.4 \times 10^{-4} \sqrt{\frac{G \mu}{10^{-10}}} & {\rm Olum\ et\ al.\ \cite{Olum:2006at}} \\
7.5 \times 10^{-4} \left( \frac{G \mu}{10^{-10}} \right)^{0.6} & {\rm Ringeval\ et\ al.\ \cite{Ringeval:2005}}
\end{array}
\right.
\ee
which shows the dominance of the loops.
The evaluation of the lensing probability and rate involves the same
integrals which depend on the gravitational wave damping cutoff in
the same way. When strings have low tension {\it loops, not horizon crossing
strings, are the favored lensing candidates.}

The appendix treats general powerlaw forms for $f$ with upper and
lower cutoffs ($\alpha_U$, $\alpha_L$) and various slopes ($\beta$)
and, most significantly, various $G \mu$ and radiation loss rates
$\Gamma_R$.  It gives approximate expressions valid for loops from
both radiation and matter epochs. It compares the simple analytic
approximations to more realistic $\Lambda$-CDM cosmologies. It
provides a range of numerical solutions illustrating how the loop
distribution varies with $\mu$, $\Gamma_R$, $\beta$, $\alpha_L$ and
$\alpha_U$.

As mentioned earlier, cosmic superstrings have different properties
from cosmic strings. The intercommutation probability of vortices is
known to be around unity, $P \simeq 1$, while that of
superstrings is rather complicated, but $P \sim g_{s}^{2}$
\cite{Jackson:2004zg}, where the string coupling $g_{s} \sim 1/10$.
Also, cosmic superstrings come in a variety of tensions and charges so
that a number of species are present in the network \cite{Tye:2005fn}.
These effects likely increase the energy density in the superstring
network compared to its cosmic strings counterpart roughly \be
\Omega_{s} \rightarrow \frac{n}{P} \Omega_{s} \ee where $n$ is the
effective number of types, $n \sim 5$. For very small $P$, it has been
argued that ${1}/{P} \rightarrow {1}/{P^{2/3}}$
\cite{Avgoustidis:2005nv}.  The implication is that the number density
of small cosmic superstring loops will be boosted with respect to
cosmic strings.

\section{Microlensing}

Microlensing refers to brightness variations of a background source
caused by a changing gravitational field somewhere along
the line of sight. The
field may be generated by a dark point-like object and the
astrophysically anticipated candidates include dim stars, white
dwarfs, neutron stars, and black holes. The paths of photons emitted by the
source are bent so that the angular area of the source visible to the
observer changes. In short, the flux from the source changes. The key point
is that one does not have to resolve the source or the lens to observe
the change. Lensing distorts a single source image or, if the impact
parameter of the photon is within the Einstein radius, lensing creates
multiple images. The area within the Einstein radius is quite small so
its much more common for magnification of a single image to occur than
the creation of multiple images. If source, lens and observer have
constant velocities, the time-dependent magnification has an {\it a priori}
known functional form. Previous surveys (MACHO, OGLE, EROS, etc.) have
searched for and identified numerous events with the expected
time-dependent achromatic form.

Now, instead of the normal astrophysical lensing candidates,
consider a stationary straight infinite string oriented perpendicular
to the observer's line of sight with respect to 
a background source.  Let
two photons from the source travel toward the string.  The photons do
not suffer any relative deflection during the fly-by as long as they
pass around the string in the same sense.  Images formed from photons
are undistorted. This contrasts with the shear and distortion
produced by a point mass.

However, there is a small angular region just like the Einstein radius
that yields multiple images as long as the source itself is small in
angular size compared to the Einstein radius. In essence, some of the
photons pass around the the string in a clockwise sense and others do
so in a counterclockwise sense. Two paths from the source to the
observer yield two undistorted images.

Sources that lie behind the string and within the characteristic
angle $\sim 8 \pi G \mu$ will appear as double images.
Unresolved, lensed sources will appear to fluctuate
in brightness by a factor of 2 as the angular region associated with
the string passes across the observer-source line of sight. The
characteristic Einstein angle is
\ba
\Theta_{E} & = & 8 \pi G \mu \nonumber\\
& = & 1.04 \times 10^{-3} \left( G \mu \over 2 \times 10^{-10} \right) {\rm arcsec}.
\ea
The characteristic angular size of a 
stellar source at distance $R$ is $\Theta_\odot = R_\odot / R$. The
relative size is
\be
{\Theta_\odot \over \Theta_E} =
4.5 \times 10^{-5} \left( 2 \times 10^{-10} \over G \mu \right) 
\left( 100 {\rm kpc} \over R \right)
\ee
which shows when the stellar source will generally be well described
as a point source.  The relativistically moving and oscillating string
will create brightness fluctuations in the background star that can be
searched for in a microlensing experiment.

The actual situation is somewhat more complicated. For a loop, as
opposed to a straight string, one expects lensing like that of a point
mass for photons with impact parameter large compared to the size of
the loop and lensing like that of a straight string for paths that
pass close to a segment of the string. We will eschew the
complications associated with small scale structure on the string and
concentrate on photons that pass close to a smooth segment of the
loop.

The characteristic scale of the smallest loops today is
\be
l_g = \Gamma_R G \mu t_{today} = 41 {\rm pc} \left( \Gamma_R G \mu \over 10^{-8}
\right) \left( t_{today} \over 13.5 {\rm Gyr} \right) .
\ee
and the characteristic mass scale associated with such a loop is
\be
M_g = 1.7 \times 10^5 \msun \left(\frac{G \mu}{2 \times 10^{-10}} \right)^2
\left(\frac{\Gamma_R}{50} \right)
\ee
both of which are small compared to Galactic scales.

For comparison, the characteristic scale of
the loops formed at equipartition is
\be
l_{max,eq}= \alpha_U t_{eq} = \alpha_U 14 {\rm kpc} \left( t_{eq} \over 4.7 \times 10^4 {\rm yrs} \right) .
\ee
Galactic scales are $R \sim 1-100$ kpc and microlensing can
probe the full range of loops generated during the
radiation era plus the small end of the loops generated
during the matter era. All this assumes $\alpha_U$ order unity.
Generally, the internal velocities associated with loops are
relativistic.

We want to answer two questions: What is the probability for lensing a
single source at distance R by a distribution of loops at a given
instant? How does the probability grow with time?

Consider a small loop of size $l$ at distance $r$. It lenses an
angular area $\Omega_L \sim (\theta_E r) l/r^2$. The probability that a single
background source at distance $R$ is lensed is the ratio of the
lensed angular area to the observed angular area in the direction
of the source. We find
\be
P_L = \int r^2 dr \int {\cal F} {dN \over dV dl} {\theta_E l \over r} dl
\ee
and ${\cal F}$ is the overdensity of loops in the Galaxy (${\cal F}=1$ gives
the lensing probability in a uniform Universe. We will estimate ${\cal F}$
below.)
Assuming the ordering $l_g < l_{max,eq} < R < \alpha_U t_{today}$ we find
for a homogeneous loop distribution
\ba
P_L & = & 3.4 \times 10^{-15} {\cal F}
\left( G \mu \over 2 \times 10^{-10} \right)
\left( R \over 100 {\rm kpc} \right)^2
\left( 1.35 \times 10^{10} {\rm yrs} \over t_{today} \right)^2 H(x,y)
\ea
where $H$ is the first moment of the loop
distribution scaled to fiducial parameters. 
This approximate analytic result is based on two joined powerlaws
for radiation and matter eras. In the appendix we show
\ba
H(x,y) & = & \left( 0.63 + 0.37 \sqrt{ y } + 0.04  \log x  \right) \\
x & = & 
\left( \alpha_U \over 0.3 \right)
\left( 10^{-8} \over \Gamma_R G \mu \right) \\
y & = & x
\left( t_{eq} \over 4.7 \times 10^4 {\rm yrs} \right)
\left( 1.35 \times 10^{10} {\rm yrs} \over t_{today} \right)
\ea
Note that as $\Gamma_R G \mu \to 0$
that $H$ varies. In particular, 
$x \propto y \propto 1/\Gamma_R G \mu$ and $P_L \propto
\sqrt{\mu / \Gamma_R}$. Extending the loop distribution to smaller
sizes ($\mu \to 0$), therefore, does not overcome the effect of
the decrease in Einstein radius.

A numerical evaluation of $P_L$ in the $\Lambda$-CDM model is
presented in Figure \ref{fig:PL-lambdaCDM} for the basic parameter
space in $G \mu-\Gamma_R$ we will consider.  The string network
has $\alpha_U=0.3$, $\alpha_L=10^{-4}$, $\beta=1.6$ with
normalization set by agreement with simulations $\Upsilon=43.6$ (see
Appendix).  \FIGURE{ \epsfig{file=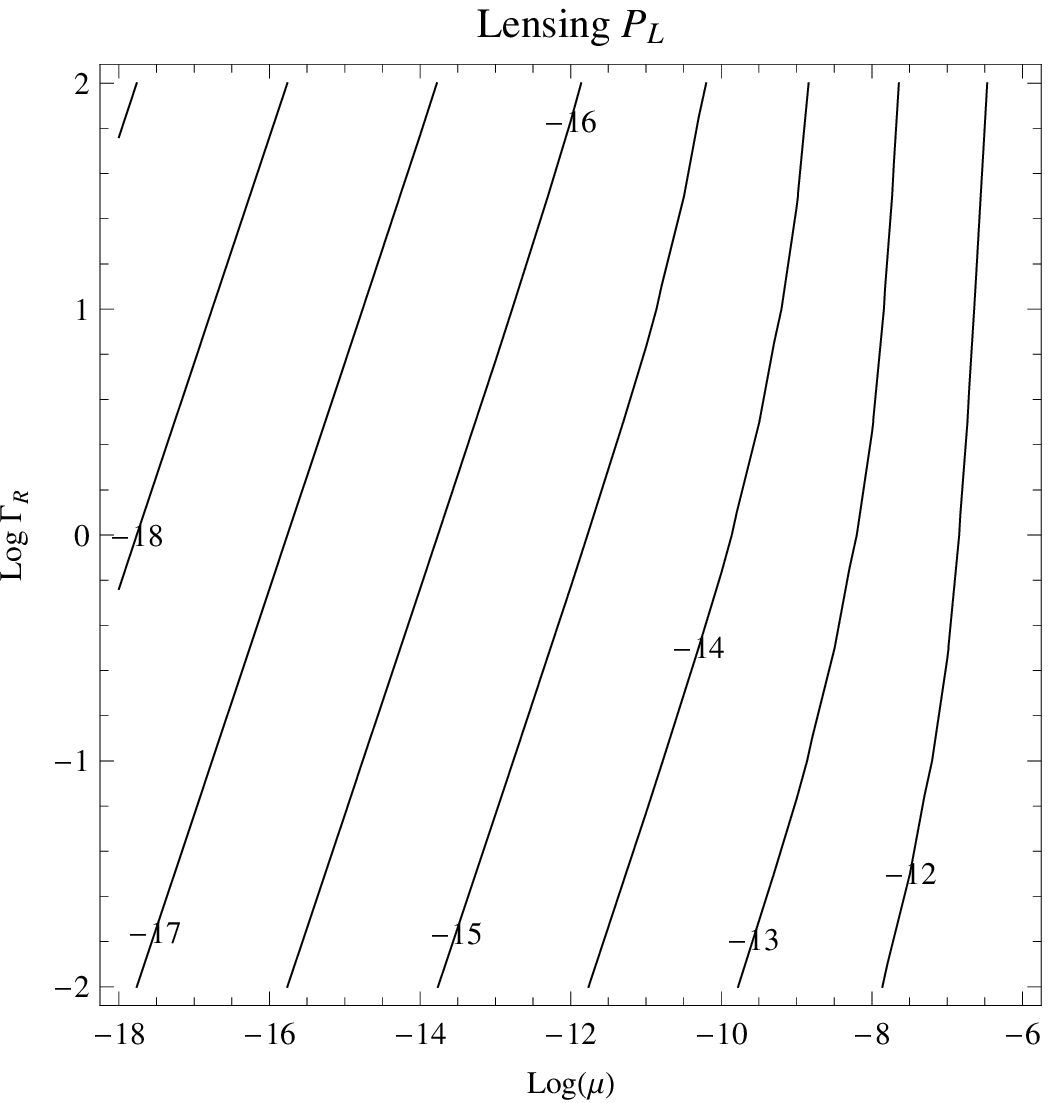}
\caption{Probability that a line of sight to a source at $R=100$ kpc is
microlensed by intervening string loops in a homogeneous Universe 
(${\cal F} = 1$).}
\label{fig:PL-lambdaCDM}
}
Here and elsewhere the numerical results we present are based on the
fiducial $\Lambda$-CDM cosmology; the approximate analytic results
illustrate the basic scalings.

Now consider an experiment that stares at a given source and looks for
the doubling in brightness on account of the passage of a loop along
the line of sight.  One loop sweeps out an area per unit time $\sim c
l/\sqrt{3}$ where the numerical factor crudely accounts for velocity
orientation effects.
The rate of change in the solid angle is ${d\Omega_L/dt}
\sim c l/\sqrt{3} r^2$. The lensing rate is
\be
R_L = \int r^2 dr \int {\cal F} {dN \over dV dl} {c l \over \sqrt{3} r^2} dl
\ee
which yields the lensing rate per source per year for a homogeneous
loop distribution
\ba
R_L & = & 2.3 \times 10^{-12} {\cal F}
\left( R \over 100 {\rm kpc} \right)
\left( 1.35 \times 10^{10} {\rm yrs} \over t_{today} \right)^2 H(x,y) .
\ea
By contrast to the situation for $P_L$, when $\Gamma_R G \mu \to 0$ we
find $R_L \propto \sqrt{1 / \Gamma_R \mu}$. 

The numerical evaluation of $R_L$ for the same $\Lambda$-CDM model
as described above is shown in Figure \ref{fig:RL-lambdaCDM}.
\FIGURE{
\epsfig{file=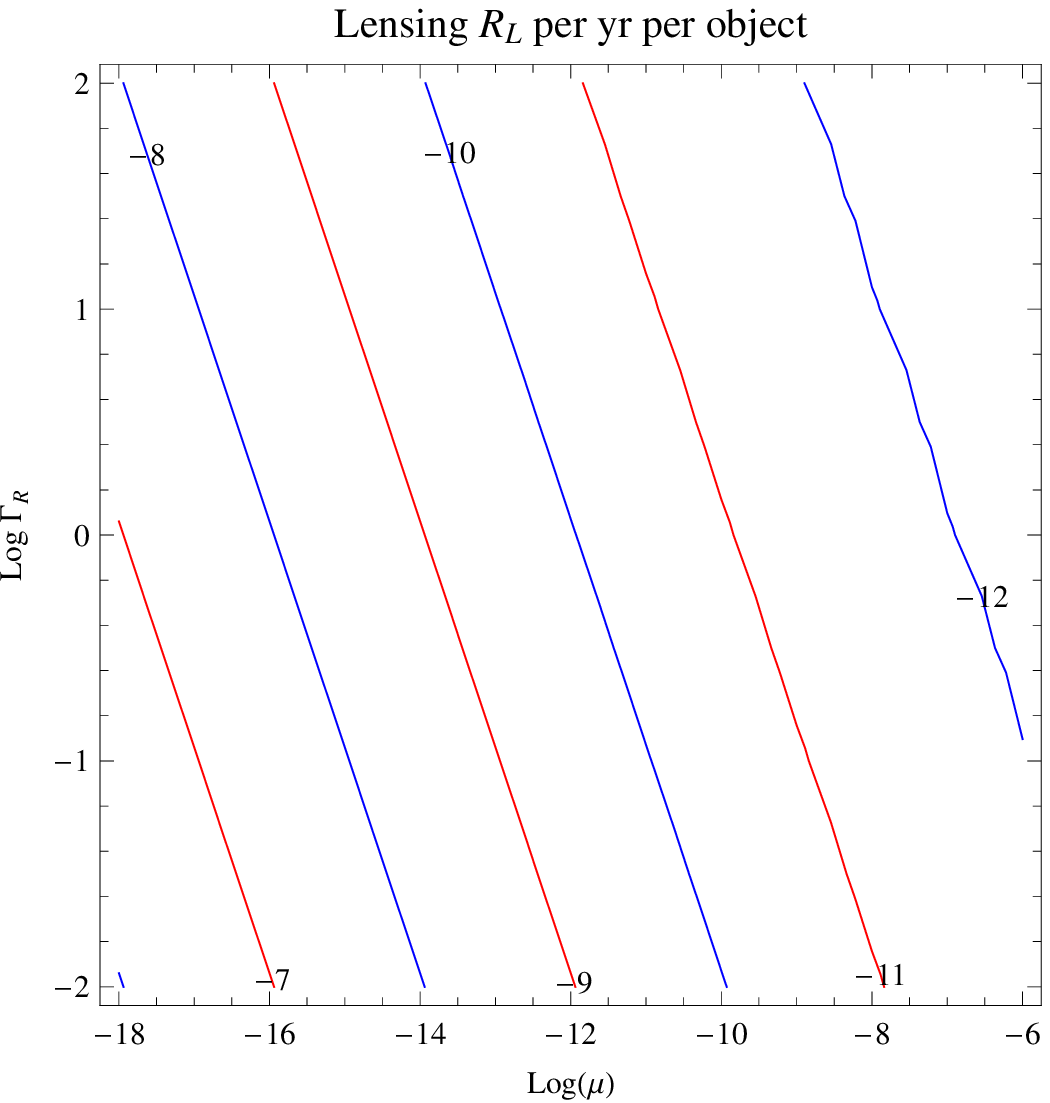}
\caption{Rate of microlensing along a line of sight to a source at $R=100$ kpc
by intervening string loops in a homogeneous Universe (${\cal F}=1$).}
\label{fig:RL-lambdaCDM}
}

We have estimated the rate $R_L$ assuming that the loop is
moving relativistically. This is generally the case for the internal
motions of the loop about its center of mass. The velocity of
the center of mass of a loop accreted to the Galaxy will be
the halo velocity $v_h$ as we describe in more detail in the
next section. New sources are lensed at a rate $(v_h/c) R_L$ with
repetition $\sim c/v_h$.

\section{Cosmology of Small $G \mu$ Loops}

The {\it center of mass velocity} of string loops within the galaxy is
a key parameter in characterizing the lensing rate for two
reasons. First, if the velocity is less than the characteristic escape
velocity from the Galaxy $v_h$ then loops will accrete and be bound to
the halo. Second, if the velocity of small loops is much less than $c$
then a single source is lensed multiple times.

The initial loop velocity is determined by the interactions in the
string network. Intercommutations between string segments can generate
relativistic center of mass motions for the newly formed loops. Our
treatment of the loop distribution is based on the assumption that all
intercommutations occurs shortly after a horizon-crossing loop is
chopped up. General scaling arguments suggest that the interaction
rate diminishes rapidly. Without interactions, the center of mass
velocity suffers cosmic drag.\footnote{We assume that the recoil
from emitted gravitational wave radiation is small.}

Assume that the initial center
of mass velocity is $v_{ci}= 0.1 c$ at time $t_i$ when the loop is born.
At later time $t$ the center of mass velocity is \cite{AlbrechtTurok89}
\ba
v_c(t) & = & \frac{v_{ci} x}{\sqrt{1 - v_{ci}^2  + v_{ci}^2 x^2}} \\
x & = & \frac{a(t_i)}{a(t)}
\ea
Figure \ref{fig:psfracv2} shows the
fraction $f_{slow}$ of the first moment of $dN/dVdl$
attributed to slow moving objects.
``Slow'' means the loop's center of mass velocity today $< 300$ km s$^{-1}$ in
the $\Lambda$-CDM model with the fiducial parameters.
\FIGURE{
\epsfig{file=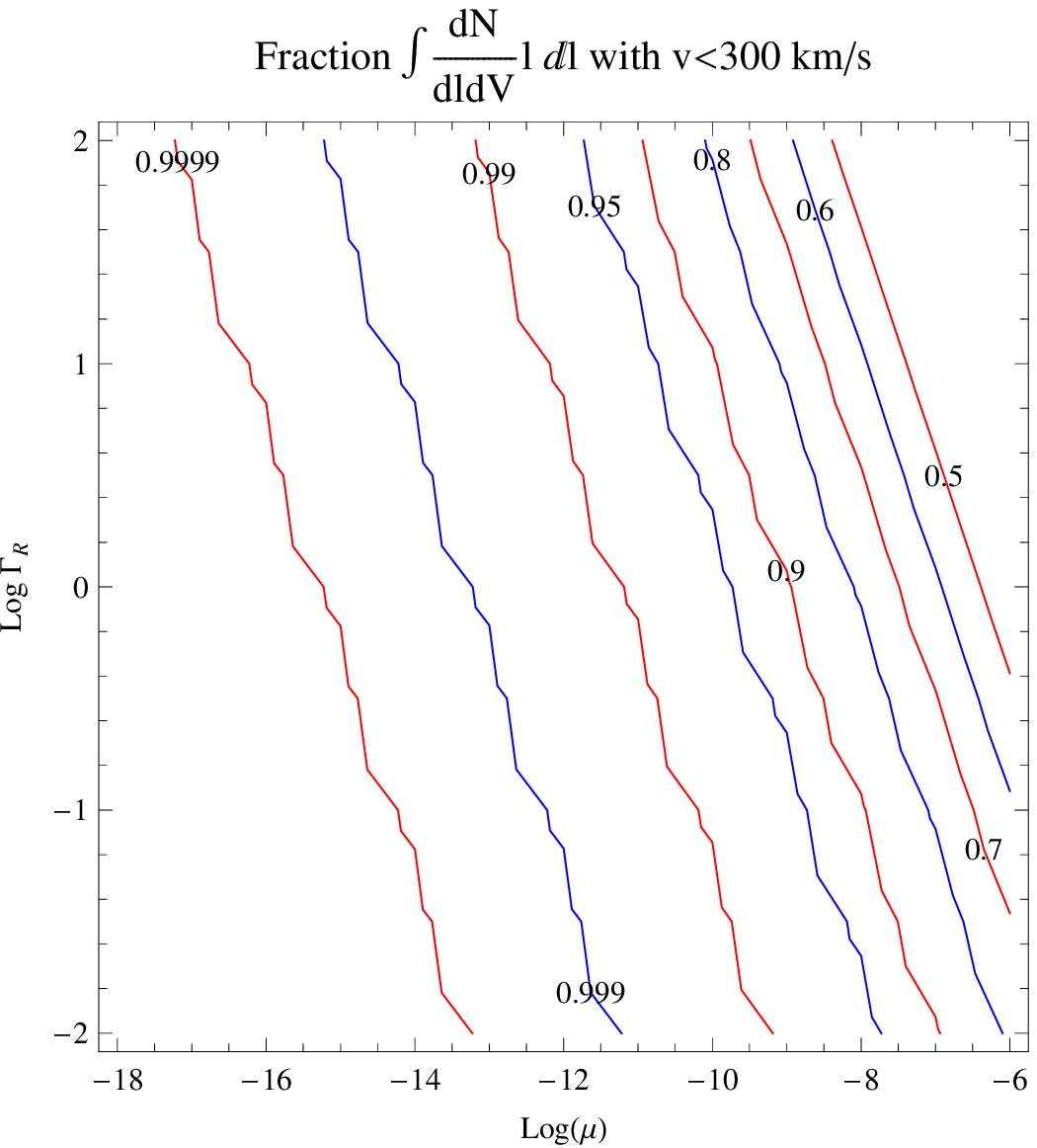}
\caption{The fraction of the length-weighted loop density distribution
with center of mass velocity $< 300$ km $^{-1}$.}
\label{fig:psfracv2}
} 
The first moment is dominated by objects moving slowly enough to
bind to the current halo. 

The scale of the gravitational potential induced by a low tension string
is small compared to that of the amplitude of perturbations entering
the horizon $\sim 10^{-5}$. At equipartition $\Omega_{loop} \sim
0.0016 \sqrt{G \mu / 2 \times 10^{-10}}$ and the mass scale of the
loop component which contributes most to $\Omega$ at that epoch is
small ($\sim 0.5 \msun \left(\frac{G \mu}{2 \times 10^{-10}} \right)^2
$). We first treat the low tension strings as ``test particles''
with respect to the baryons and CDM perturbations. As the
perturbations begin to collapse after equipartition the slow moving
loops fall into the gravitational potentials which will eventually
give rise to galaxies.

The observed overdensity of material in the Galaxy is 
\ba f_{over} & = &
\frac{\rho_{Gal}}{\Omega_M \rho_{crit}} \\ & \sim & 1.5 \times 10^5
\ea 
where we have used $\Omega_M = 0.3$, $H_0=70$ km s$^{-1}$
Mpc$^{-1}$, and a galactic mass of $2 \times 10^{11} \msun$ in a
characteristic size of $20$ kpc \cite{FichTremaine91}. 
The size scale has been selected in consideration of
the location of stars of interest for microlensing.
An estimate for the local enhancement of loops is
\be 
{\cal F} = f_{slow} f_{over} 
\ee
i.e. roughly the overdensity of the Galaxy itself. 

The above estimate for ${\cal F}$ is a lower bound because the actual
dynamical evolution of the loops is not dissipationless.  At
equipartition, a loop begins to accrete matter.  A novel aspect of the
structure formation in this scenario is that suitably small loops will
radiate and disappear but leave behind bound sub-galactic clumps of
matter. Such objects are of considerable interest in their own right
but here we focus on the mass density of the smallest surviving loops
of mass $M_g$ in today's Galaxy. Such a loop accretes a mass $\sim
(1+z_{eq}) M_g$ with $z_{eq} \sim 3570$ (in the approximation of a
flat, matter-dominated cosmology). This bound, composite object lives
in the larger-scale growing galactic potential and it will be dragged
by dynamical friction towards the center. As an example, consider the
case for $G \mu = 2 \times 10^{-10}$, where $M_g \sim 1.7 \times 10^5
\msun$ and clump mass $\sim 6.4 \times 10^8 \msun$. Let the clump move
on a circular orbit in an isothermal halo with rotation velocity $220$
km s$^{-1}$.  The radial drift takes it from $\sim 36$ kpc to $20$ kpc
over a Hubble time \cite{BinneyTremaine87}.  The net motion with
respect to the halo increases the density of such loops within $20$
kpc by at least an additional factor $\sim 1.8$. The net radial drift
for a loop of length $l$ scales $\propto (\mu l)^{1/2}$ and the drift
for the smallest surviving loop scales $\propto \mu \sqrt{\Gamma_R}$.
These dissipational effects mean that loops, like baryons, may be
over-concentrated with respect to the dark matter.  The effect on the
smallest loops is important if $G \mu \simgreat 10^{-10}$. The
distribution of loop size will also be tilted within the Galaxy
compared to $dN/dVdl$ on account of the dissipative processes.

We make a conservative estimate for $\cal F$ in our numerical
calculations by ignoring these dissipative enhancements to the loop
density.

\section{Practical Lensing}

There are a number of characteristic timescales relevant to
experimental detection of lensing. For the lensing itself,
the characteristic duration of the event is
\ba
\delta t_{lens} & = & {R \theta_E \over c} \nonumber\\
         & = & {R 8 \pi G \mu \over c} \nonumber \\
         & = & 6.3 \times 10^3 {\rm sec}
\left( R \over 100 {\rm kpc} \right)
\left( G \mu \over 2 \times 10^{-10} \right)
\ea
The characteristic time for the (smallest) loop to oscillate is
\be
t_{osc} \sim {l_g \over c} \sim 135 {\rm yrs} \left( {\Gamma_R G \mu \over 10^{-8}} \right)
\ee
and this governs the repetition timescale.

The observational timescales are $\delta t_1$, the time for an individual
observation, $\Delta T$, the duration of the experiment, and $\delta
\Delta T=\Delta T/N_{rep}$, the characteristic interval between
observations where $N_{rep}$ is the number of times a star is
visited over the course of the experiment.

To make this concrete, we will consider the Gaia mission which will
monitor $N_*=10^9$ stars for $\Delta T = 5$ yrs with
fluxes exceeding a mission-defined broad band magnitude $G=20$. 
On average, the Broad-band photometer
will observe each star $N_{rep} = 80$ times
but the interval between
observations is not fixed; it will vary from $\sim 30$ minutes to
$\Delta T/N_{rep}$. An individual
observation is $\delta t_1 \sim 3.3$ seconds.  Let $\delta \Delta T
\sim \Delta T_{obs}/N_{rep}$ be the characteristic interval between
repeat observations. The mission observes $\sim 10^3$ stars at a time
(note $N_* N_{rep} \delta t_1 >> \Delta T$); all astrometric and
photometric results for
individual stars are derived by detailed analysis of the joint
observations.

The accuracy of the photometry depends upon the brightness of the
source. The limiting magnitude $G=20$ corresponds to objects with a
range of usual visual magnitudes $V \sim 20-25$. Individual
observations for sources with $V=20$, $21$, $22$ and $23$, for
example, have relative flux accuracies $2.5$\%, $5$\%, $10$\% and $26$\%
respectively and are more than adequate to see a factor of 2 change
due to microlensing at the limiting magnitude.

A star like the Sun (type G2V) will be visible to a limiting magnitude
$V=20.2$ or a distance approximately $12$ kpc.
To make estimates of the number of lensing events that Gaia is
capable of observing we need to account for the distribution
of stars and the ``efficiencies'' 
with which detections can be made.  We assume here that
the stars observed by Gaia are uniformly distributed in
space and concentrate on the detection efficiency
\be
f_{det} = f_{mag} f_{size} f_{time}
\ee
where $f_{mag}$ is the flux limit, 
$f_{size}$ is the source size cutoff
(so that the angular size $\Theta < \Theta_E$),
and $f_{time}$ accounts for sampling in time. We adopt a crude
model
\be
f_{time} =
\left\{
\begin{array}{cc}
0        & {\rm if\ \ } \delta t_{lens} < \delta t_1 \\
\frac{\delta t_{lens}}{\delta \Delta T} & {\rm if\ \ } \delta t_1 < \delta t_{lens} < \delta \Delta T \\
1        & {\rm if\ \ } \delta \Delta T < \delta t_{lens} < \Delta T  \\
0        & {\rm if\ \ } \Delta T < \delta t_{lens}
\end{array}
\right.
\ee
Any lensing event which is shorter than the length of a single
observation or longer than the length of the entire experiment cannot
be detected; all remaining events longer than the average interval
between observations will be seen but only a fraction of events
shorter than it will be. The assumption is that the observational
intervals are uniformly spread between the shortest and longest
periods. A more accurate description of the schedule of observations
will readily improve the model for $f_{time}$. To find the detectable
lensing rate, we recalculate $R_L$ with $f_{det}$ to give
$R_L^{Gaia}$. The expected number of events is $N^{Gaia} = R_L^{Gaia}
N_* \Delta T$.

The expected number of lensings detectable by Gaia are shown in
Figure \ref{fig:psNgaia2-lambdaCDM} by red/blue contours. These include all
efficiency factors (magnitude, size and time) with each contour
labeled by the log of the number of detectable lensings. Note that the
expected number of detectable events decreases as $G \mu \to 0$ at fixed $\Gamma_R$ (for $G
\mu < 10^{-8}$), a consequence of decreased detection efficiency for
short duration events ($f_{time} \to 0$). The sharp cutoff as $G \mu
\sim 10^{-13}$ is where the angular size of the star becomes comparable
to the Einstein angle ($f_{size} \to 0$).  The number of events
depends on the path length (set by the flux limit $f_{mag}$) and the
intrinsic loop density.

To illustrate the important effect of size and time cutoffs, we also
display the expected number of lensing events if detections were
limited only by flux considerations (green dashed contours). Note that
the expected numbers increase as $G \mu \to 0$. These curves give
some guidance on how alternative experiments might fare by adjusting
$N_{rep}$, $N_*$ and $\Delta T$.

\FIGURE{
\epsfig{file=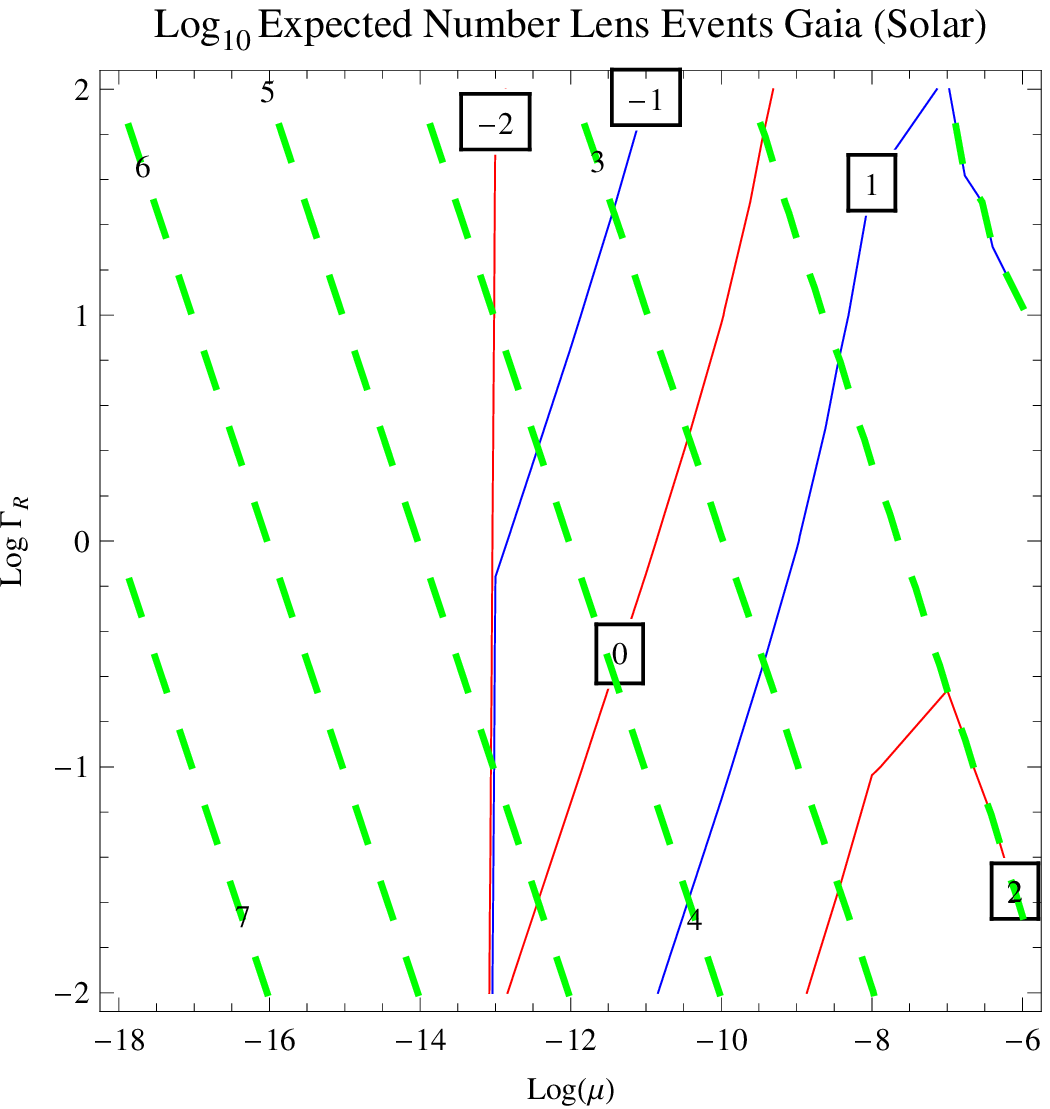}
\caption{Log of the number of microlensing events detectable by a Gaia-like
mission looking at $10^9$ sources over $5$ years. In this plot
all sources are assumed to be solar mass stars (G2V) within
$12$ kpc. The string loop density is based on the fiducial
powerlaw model evolved in a $\Lambda$-CDM cosmology. The local
string overdensity is
${\cal F} = f_{slow} f_{over}$ as described in the text. The
red/blue lines include magnitude, time and angular efficiency
factors; the dashed green lines include only the flux cutoff.}
\label{fig:psNgaia2-lambdaCDM}
}
An experiment with duration $\Delta T < t_{osc}$ will see at most one
lensing event. Conversely, if $\Delta T > t_{osc}$ the experiment has
the capability of seeing multiple lensing events from the same source.
If $G \mu < 7 \times 10^{-12}$ multiple lensing may be observed
by Gaia. In any case, a potential lens can be re-observed by
other means even after the mission has ended.

The figure has been constructed for a ``typical'' star within the
Galaxy.  The objects included in Gaia's catalog will depend on many
factors -- stellar type, interstellar obscuration, crowding, etc. --
and these will alter the effective cutoffs and the number of
events. For example, bright halo giants will be visible within
approximately $41$ kpc (type G0III; radius $\sim 3.3
R_\odot$). Compared to the more numerous G2V class, the lensing rate
per star should be $\sim 14$ times larger on account of longer
pathlengths and longer lensing durations. On the other hand, the
cutoff at small $G \mu$ will be comparable to G2V because the angular
sizes are comparable at the limiting distance. 

At the flux limit, a star's angular size $\propto 1/T_{eff}^2$ where
$T_{eff}$ is the star's effective temperature. Hotter stars can be
seen further and have smaller angular size at the limiting distance.
Consequently, the condition that the Einstein angle exceeds the
stellar angular size implies a minimum tension detectable by
an optical microlensing experiment $\propto
1/T_{eff}^2$. The inclusion of stars hotter than the sun may provide
sensitivity to tensions beyond the cutoff indicated in the figure. Of
course, the number of such stars is an important consideration: hot
white dwarfs, for example, would probe very low tensions but are not
visible to large distances.  Likewise, the duration of the lensing
event should exceed the timescale of a single observation to measure
the doubling in image brightness. At the flux limit, the minimum
tension $\propto 1/L^{1/2}$ where $L$ is the stellar luminosity. The
inclusion of stars more luminous than the sun may provide sensitivity
beyond the cutoff. An accurate assessment will depend on the make up
of Gaia's catalog, a subject to which we hope to return.

The Broad-band photometer measures $5$ colors during an observation
but this information is insufficient to allow identification and
classification of stellar types throughout the HR diagram; it will be
supplemented by photometry in $14$ other bands from the Medium-band
photometer, a distinct instrument on board. The mission expects to
observe $\sim 18 \times 10^6$ variable stars \cite{Gaia} and the
extensive color coverage is essential to facilitate identification and
classification of known variables types (pulsating stars, eruptive
variables, etc.). This circumstance will prove a great benefit if
string microlensing is sought. String microlensing is almost
achromatic (the Kaiser Stebbins effect gives relative frequency shifts
order $\sim G \mu \gamma (v/c)$; since $G \mu$ is small the shift is
small for mildly relativistic loop motion) whereas variable stars
typically show color changes. The key point is that a cosmic string
microlensing event should be an {\it achromatic} change in brightness
by a factor of $2$.  The situation is qualitatively similar to that
faced by ground-based microlensing experiments which anticipate an
achromatic change in an {\it a priori} known form for the time-dependent
amplification factor.

Finally, we note that the Broad and Medium-band photometers have their
own time sampling characteristics. Depending upon the detailed
scheduling of scans, there may be an improved detection
efficiency compared to our simple estimate.

\section{Remarks} 

When string tension is low the density in loops exceeds that in long
strings because the timescale for
gravitational damping becomes long. The total string energy density
today is dominated by the loops originally formed during the radiation
era. This preponderance of loops over long strings motivates
consideration of the cosmology of loops. The evolution of these
objects turns out to be completely distinct from that of the long
strings originally studied with $G \mu \simgreat 10^{-6}$.

The loops damp by cosmic drag and are utterly inconsequential in terms
of structure formation on scales much larger than the
galaxy. Instead, they fall into
the potential wells created by cold dark matter and baryons after
equipartition. The Galactic halo number density of loops is enhanced
over the universe's average value by at least
the Galaxy's overdensity $\sim 10^5$. Dissipational effects which
depend upon the size of $G \mu$ may
further increase the overdensity.

The loops within the Galaxy can be observed by microlensing using
stars as sources. To the extent that stars are point-like and bright
low tension strings can be observed. Simple estimates of a
microlensing survey based on the capabilities of the Gaia mission
suggest that many such events may be detected. If loops are moving at halo
velocities the lensing of a given source should repeat $\sim 10^3$
times.

Detailed observations of such a lensing source will have much to tell
us about string tension and the number density of loops.

We recognize many gaps in this general story
\begin{itemize}
\item{What processes are responsible for producing the ${\cal F}=1$ 
(average, unclustered)
distribution of loop sizes? When does the rate of
interactions between loops and/or long strings become
negligible? When does cosmic drag predominate over velocities
induced by intercommutation?}
\item{How do loop-loop interactions develop as the
galaxy grows? Does intercommutation end up ejecting loops
from the galaxy? How does the distribution
function of loop sizes change on account of the intercommutations
in the dense environment? How small are the smallest loops?
Is the gravitational lifetime
significantly altered? Can one use this to measure
or learn about $p$? The loop size distribution is
potentially measurable via microlensing and locally generated
gravitational waves.}
\item{What are the astrophysical constraints on massive loops
moving around galaxy? How are loops distributed within the
Galaxy?}
\item{What string and loop parameters can we deduce by observing
repetitive lensing? Does small scale structure on the loops increase the
microlensing repetition rate?}
\item{Incorporate lensing over radii ranging from larger than
the loop size to smaller. Is the Kaiser-Stebbins effect
detectable from a known lens?}
\item{An accurate calculation of Gaia's capabilities for
string microlensing requires
in depth analysis of the time-based observing strategy for both
the broad and medium band photometers.}
\item{Can one design an experiment tailored to doing a better job
for detecting microlensing than a mission like Gaia? It is worth noting that
the Gaia sample includes photometry
much more accurate than what is needed to detect microlensing.
Could one generate a larger catalog
with more poorly determined flux measurements still adequate to
detect string microlensing? Would there be enough information
to rule out variable stars?
}
\item{How can one optimize the choice of $\Delta T$, $N_{rep}$ and
$\delta t_1$ to investigate particular ranges of $G \mu$?}
\end{itemize}

Detecting cosmic superstrings and measuring their tension can reveal
fundamental information about superstring theory and elucidate the
large-scale contents as well as the remote evolutionary history of our
own universe. An observable, nearby source of strings to study promises
to advance, qualitatively and quantitatively, these goals.

\acknowledgments
We thank Joe Polchinski, Hassan Firouzjahi, Louis Leblond, Irit Maor,
Gary Shiu, Ira Wasserman, Mark Wyman, Ben Shlaer, Ken Olum, Xavier
Siemens, Konrad Kuijken, Tanmay Vachaspati
for useful discussions. The work of DFC is
supported by the National Science Foundation (AST-0406635) and
by NASA (NNG05GF79G). The work of SHT is
supported by the National Science Foundation under grant PHY-0355005.

\appendix
\section{Appendix}

We start with a form generally consistent with the assumption
of a scaling network
\be
\frac{dN(l,t)}{dl dt dV} = t^{-5} f(x) \qquad \text{with $x=l/t$}
\ee
where $l$ and $V$ are measured in physical (not comoving) lengths.  We
make the assumption that the chopping of long strings all the way down
to the smallest loops occurs in a time short compared to the expansion
timescale. Of course, on the Hubble scale this won't be a very good
approximation but it should get better at smaller scales and thats
what is of most interest here. Loop length $l$ is the time-averaged
physical length associated with the loop when the loop
itself is small compared to the scale of the horizon. A cosmological
simulation is required to establish the form for $f$ which describes
loops with sizes comparable to the scale of the horizon.

Motivated by the numerical simulation in
Ref.\cite{Vanchurin:2005yb,Vanchurin:2005pa}, we take the loop
production to be given by a power law distribution with upper
and lower cutoffs
\be
f(x) = A x^{-\beta} \qquad \text{for $\alpha_L < x < \alpha_U$}
\ee
and zero otherwise. In addition to the upper limit, we impose a lower
limit on the the range of loops. This is a manifestation of the
cutting up process and not due to gravitational wave emission. In this
approximate description, dynamical processes chop up strings to give
the loop spectrum instantly. Gravitational radiation acts once
the spectrum has been created.

The scaling network is characterized by some
inter-string distance $d(t) =\gamma_{s} t$, defined so that the density in
long strings is $\rho_\infty =\mu/d^2$.  Conservation of energy 
then gives 
\be 
\int \mu l \frac{dN}{dl dV dt} dl = - \frac{d \rho_\infty}{dt}
\ee
or
\be
\int_{\alpha_L}^{\alpha_U} x f(x) dx = 2 \frac{1}{\gamma_{s}^2}\left(1-\langle
v^2\rangle\right)
\ee
where $\langle v^2\rangle$ is the square of string velocity averaged
along the length of long strings. 

The energy balance between long strings and the loop distribution
gives 
\be 
A = \frac{2 \left( 1 - \langle v^2 \rangle \right) \left( 2 - \beta
\right)}{\gamma_s^2 \alpha_U^{2-\beta} \left( 1 - \left( \alpha_L
/\alpha_U \right)^{2 - \beta} \right)}
\ee 
Numerical simulations \cite{Vanchurin:2005pa} derived $A = 82 \pm 2$,
$\alpha_L = 10^{-4}$, $\alpha_U = 10^{-1}$ and $\beta = 1.63 \pm 0.03$.
From these we deduce the numerical value for
\be
\Upsilon \equiv \frac{1- \langle v^2 \rangle}{\gamma_s^2} \simeq 43.6
\ee
which we will hold fixed even as we consider models with different
characteristic values of $G \mu$, $\alpha_L$, $\alpha_U$, $\beta$ and different
expansion dynamics $n \ne 2/3$).

It is generally thought that a loop of length $l$ decays by
gravitational radiation in time $l/(\Gamma G\mu)$, where $\Gamma$ is
the dimensionless decay rate of order 50. We write our model with
two generalizations of the canonical description.
\begin{itemize}
\item{$\Gamma_R$:} We introduce a distinct damping rate for loops,
$\Gamma_R$. It is possible that the average loop damping rate
differs substantially from $\Gamma$ 
because small loops, large loops and horizon-crossing
strings intercommutate.
\item{$\tau$:} Lifetime in the network might be a {\it nonlinear}
function of $l$. This could arise because of
the nontrivial interaction of intercommutation and damping.
\end{itemize}

If lifetime is linear in length then
loops of length $l$ born at time $t'$ and observed at time $t$ must
have
\be
l > \Gamma_R G \mu (t - t') \qquad \text{for $t>t'$}
\ee
Our model assumes that loops are unaffected by gravitational wave
damping until they reach the end of their life at which point they are
abruptly removed from the population.

The loop density at time $t_2$ due to loops produced during the
interval ${\cal I} = (t_0 ,t_1)$ is found by integrating the
production rate density production of loops:
\be
{dN \over dV dl}(t_2;{\cal I}) = {1 \over V(t_2)} \int_{t_0}^{t_1} dt' {dN \over dV dl dt'} V(t') \theta(t_2 -t' < \tau(l))
\ee
Here $V(t)$ is physical volume. Inserting the production rate
density
\be
{dN \over dV dl}(t_2;{\cal I}) = {1 \over V(t_2)} \int_{t_0^*}^{t_1^*} dt' {f({l \over t'}) V(t')\over t'^5}
\ee
where
\be
t_1^* = \min{\left( t_1,{l \over \alpha_L} \right) }
\ee
\be
t_0^* = \max{\left( t_0,t_2-\tau(l),{l \over \alpha_U} \right) } .
\ee

\subsection{Powerlaw $f$ and $a(t)$}

During ${\cal I}$ assume that the scale factor $a \propto t^n$, the volume $V \propto t^{3n}$, and $f(x) = A x^{-\beta}$ for $\alpha_L < x < \alpha_U$ (and $0$
otherwise). The loop density at $t_2$ is
\be
{dN \over dV dl}(t_2; {\cal I}) = {\cal G}
{A \over \zeta l^\beta t_2^{3n} (t_0^*)^\zeta} 
\left( 1 - \left[ t_0^* \over t_1^* \right]^{\zeta} \right)
\ee
where
\be
\zeta = 4 - 3n -\beta .
\ee
For typical values of $\beta$ and $n$ we have $\zeta > 0$. Now
the density at $t_2$ depends upon how $V$ varies beyond ${\cal I}$.
If $n$ is constant
over the entire interval $t_0$ to $t_2$ then ${\cal G} = 1$. If loops
are born in the radiation-dominated epoch ($n=n_1=1/2$), 
which extends from $t \sim 0$ to
$t_{eq} \approx 4.7 \times 10^4$ yrs, and are observed in the
matter-dominated and/or lambda-dominated epochs we have
a non-trivial $\cal G$. For the ordering $t_1 < t_{eq} < t_2$,
${\cal G} = V_+(t_{eq}) t^{3n_1}/V_+(t) t_{eq}^{3n_1}$ where $V_+$ is
the volume for $t>t_{eq}$. If $a(t) \propto t^{n_2}$ for $t>t_{eq}$ then ${\cal G}
= (t/t_{eq})^{3(n_1-n_2)}$. This reduces to ${\cal G} = (t/t_{eq})^{-1/2}$
for $n_1=1/2$ and $n_2=2/3$.

It will prove useful to rewrite the general result in the following form
\be
{dN \over dV dl} = {\cal G} {\cal N} {\alpha_U^{2-3n} \over l^4} \left( l \over t \right)^{3n} 
\left( l \over \alpha_U t_0^* \right)^\zeta
\left( 1 - \left( t_0^* \over t_1^* \right)^\zeta \right)
\ee
\be
{\cal N} = \frac {2 (2-\beta ) \Upsilon }
{\zeta  \left( 1 - {\left( \alpha_L \over \alpha_U \right)^{2 - \beta}} \right) } .
\ee

Figure \ref{fig:today-string-distribution} presents an estimate of the
loop density today obtained by adding the loops created during the
radiation and matter-dominated eras.  Specifically, we take $n=1/2$,
$t_0=0$, $t_1=t_{eq}$ followed by $n=2/3$, $t_0=t_{eq}$,
$t_1=t_2=t_{today}$.  For lack of more precise numbers we take
$\alpha_L=10^{-4}$, $\alpha_U=0.3$, $\Gamma_R G \mu = \alpha_G =
10^{-8}$ (i.e. lifetime linear in $l$: $\tau(l)=l/\alpha_G$),
$\beta=1.6$, $\Upsilon = 43.6$ during both epochs. These choices
imply ${\cal N}_1 = 40$ and ${\cal N}_2 = 91$. The background
cosmological model has $t_e = 4.7 \times 10^4$ yrs and the current
epoch $t_{today}=1.35 \times 10^{10}$ yrs.  The estimate of the number
density has a cutoff at $l \simeq 10^{-8}$ from gravitational damping;
the very slight 
kink at loop size $l \simeq 10^{-6}$ is roughly equal to $\alpha_U
t_{eq}$ i.e. of order the horizon at equipartition.

\FIGURE{
\epsfig{file=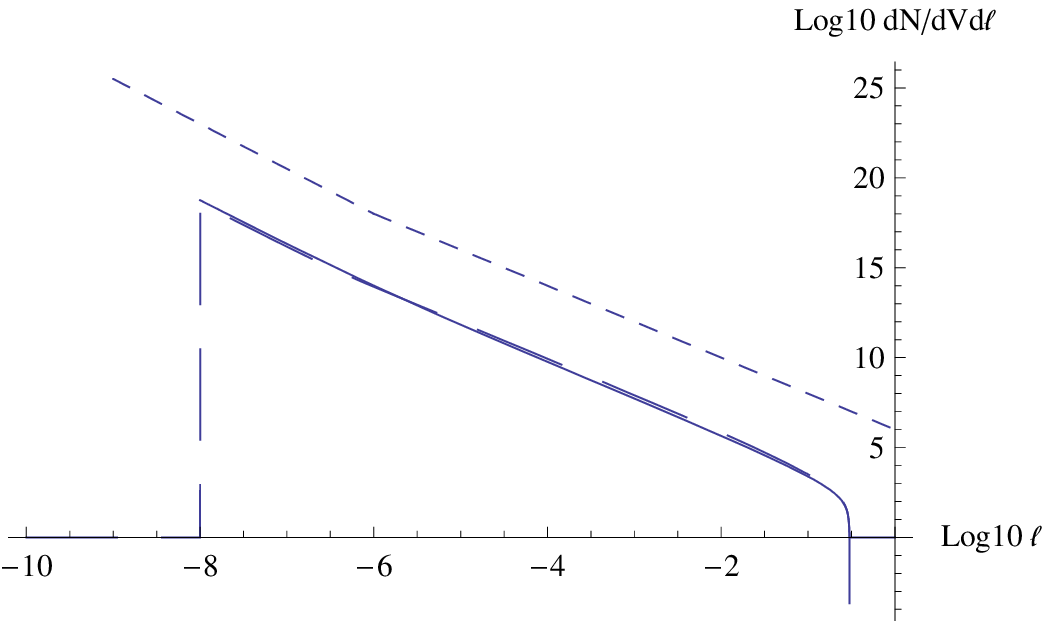}
\caption{The number density of today's string distribution based on
two powerlaw semi-analytic estimate (long dashs) and $\Lambda$CDM
model (solid line). Two straight lines (short dashs)
with slope -2.5 (radiation) and -2 (matter) are included for
comparison. }
\label{fig:today-string-distribution}
}

Two straight lines are included in the figure and are evidently close
approximations to the slope of $dN/dVdl$ over an intermediate
regime. The slope is determined by the expansion rate.  Assume a pure
powerlaw expansion with $t_0=0$ and $t_1=t_{today}$, $\alpha_L <<
\alpha_U$ and $\zeta > 0$ (e.g.  $\beta = 1.6$ and $1/2 \le n \le 2/3$
gives $0.9 \ge \zeta \ge 0.4$). Define the lengths $l_{min}=\alpha_L t$,
$l_{max}=\alpha_U t$ and $l_g = \alpha_G t$.  For $l$ large
compared to both the gravitational damping and the minimum chopping
scales $t_0^* \to {l \over \alpha_U}$ and for $l$ small compared to the
horizon $l \ll l_{max}$ we have
\be
{dN \over dV dl} = {\cal N} {\alpha_U^{2-3n} \over l^4} \left( l \over t \right)^{3n} 
\ee
and
\be
{\cal N} = \frac{2 (2-\beta ) \Upsilon }{\zeta }
\ee
This form shows that the slope of the loop distribution at
intermediate scales is completely determined by the rate of expansion:
$dN/dVdl \propto l^{-2.5}$ for $n=1/2$ and $\propto l^{-2}$ for
$n=2/3$ as previously demonstrated by Vanchurin et al. 
\cite{Vanchurin:2005pa}.

Figure \ref{fig:today-yesterday-distribution} compares the distributions at $t=t_{today}$ and at $10^{-2} t_{today}$.
The gravitational wave damping
destroys the small scale loops while the expanding horizon continuously
gives rise to larger scale loops. The expansion decreases the number density.
The intermediate form, supplemented by the appropriate $\cal G$, will
generally be sufficient to describe the loop distribution for loops
that cannot have decayed ($l > l_{g,2}$ where $l_{g,2} = l_g(t_2)$) and
smaller than the largest loop formed ($l < l_{max,1}$ where $l_{max,1}=\alpha_U t_1$), presuming the ordering $l_{g,2} < l < l_{max,1}$.
Of course, the full expression is needed
to describe the exact form near small and large $l$ and to deal with
situations in which the intermediate regime does not have enough
time to form.
\FIGURE{
\epsfig{file=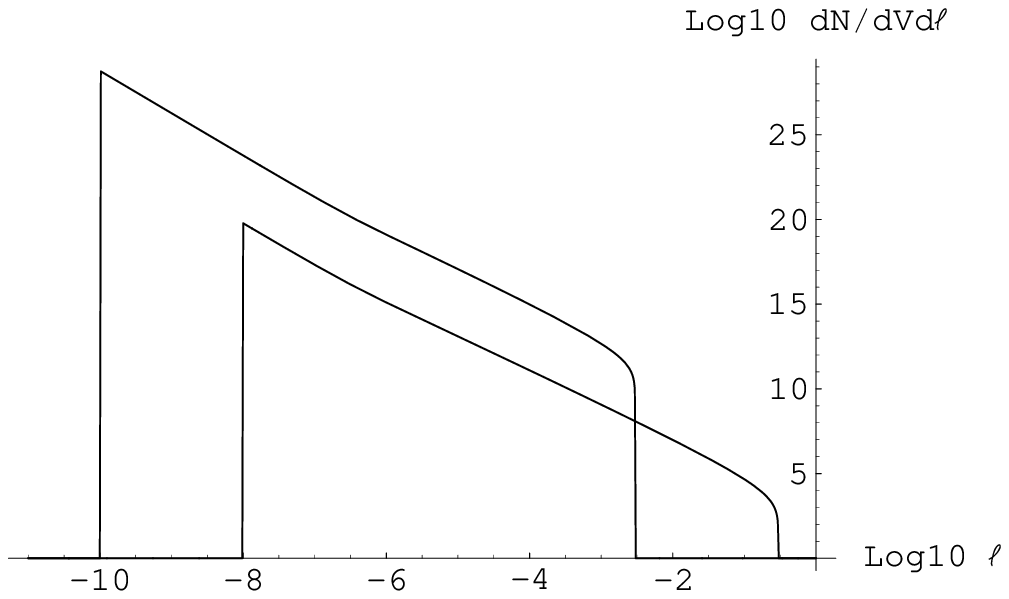}
\caption{Number density of the
string length distribution at two epochs $t_{today}$ and $10^{-2}t_{today}$
in the two powerlaw model. }
\label{fig:today-yesterday-distribution}
}

We are primarily interested in understanding
whether various integrals are dominated by the small or large scale
part of the distribution. Consider the density distribution of $l^Q$;
$Q=0$ gives number density, $Q=1$ gives length (needed for energy
density and lensing estimates), and so forth.
Using the intermediate form for $l_{g,2} < l < l_{max,1}$
\be
\int l^{Q} {dN \over dV dl}(t_2;t_0,t_1) dl \sim
{\cal G} {{\cal N} \alpha_U^{2-3n} \over t_2^{3n} (Q-3+3n)} 
\left( l_{max,1}^{Q-3+3n} - l_{g,2}^{Q-3+3n} \right)
\ee
Clearly, $Q < 3-3n$ is dominated by small scale loops ($Q<3/2$ for $n=1/2$;
$Q<1$ for $n=2/3$). Assuming
$\alpha_U t_1 \ge \alpha_G t$ the integrations give the relatively
simple forms
\be
\int l {dN \over dV dl} (t;t_0,t_1) dl 
\sim {\cal G} {{\cal N} \over t^2}
\left\{
\begin{array}{cc}
2 \left( \sqrt{ \alpha_U \over \alpha_G } - \sqrt{ t \over t_1 } \right) & {\rm if\ \ } $n=1/2$ \\
\log \left( \alpha_U t_1 \over \alpha_G t \right) & {\rm if\ \ } $n=2/3$
\end{array}
\right.
\ee

The total integral is formed from the sum of loops
created during the radiation and matter-dominated epochs 
(${\cal G}_1 =(t_{eq}/t)^{1/2}$, ${\cal G}_2 = 1$).
We assume
(for lack of any more complete information)
$\beta_1=\beta_2$, 
$\alpha_{U,1}=\alpha_{U,2}$ and 
$\alpha_{G,1}=\alpha_{G,2}$. The final result for $t=t_{today}$ is
\be
\int l {dN \over dV dl} dl 
\sim {1 \over t^2}
\left( 2 {\cal N}_1 \left(
\sqrt{ \alpha_U t_{eq} \over \alpha_G t } - 1
\right)
 + {\cal N}_2 \log \left( \alpha_U \over \alpha_G \right) \right) .
\ee
If $\alpha_G t_{today} \ll \alpha_U t_{eq}$ the loops created before $t_{eq}$ dominate and 
\be
\int l {dN \over dV dl} dl 
\to {2 {\cal N}_1 \over t^2}
\sqrt{ \alpha_U t_{eq} \over \alpha_G t } 
\ee

The total energy density is 
\be
\rho_{loop} = \mu \int l {dN \over dV dl} dl
\ee
Figure \ref{fig:today-loglength-distribution} plots $l^2 {dN/dV dl}$ so that contributions in
different logarithmic intervals of $l$ can be directly compared. Evidently,
for the parameters adopted the
small loops formed during at $t < t_e$ dominate $\rho_{loop}$. This
result (and other qualitative results alluded to above) will be
altered if $\alpha_G/\alpha_U > t_{eq}/t_{today} \sim 3.5 \times 10^{-6}$.
\FIGURE{
\epsfig{file=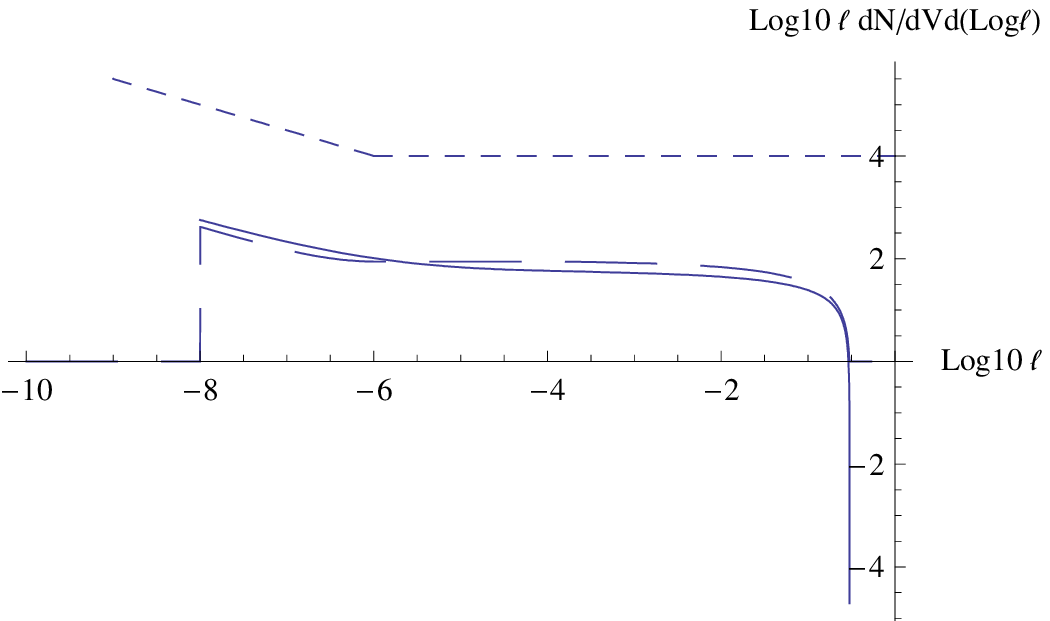}
\caption{Today's string number distribution weighted by length; lines
as in previous plot;
straight (short
dashed) lines have slope -.5 (radiation) and 0 (matter).}
\label{fig:today-loglength-distribution}
}

For a flat, matter-dominated cosmology (i.e. ignoring $\Lambda$) we
find that
\be
\Omega_{loop} = 6 \pi G \mu
\left( 2 {\cal N}_1
\left( \sqrt{ \alpha_U t_{eq} \over \alpha_G t } - 1 \right)
 + {\cal N}_2 \log \left( \alpha_U \over \alpha_G \right) \right) .
\ee

\subsection{Analytic Estimate}
Define two dimensionless parameters, scale to ``typical'' parameters, as
follows:
\ba
X & = & X_0
\left( \alpha_U \over 0.3 \right)
\left( 10^{-8} \over \Gamma_R G \mu \right) \\
X_0 & = & 3 \times 10^{7} \\
Y & = & Y_0
\left( X \over X_0 \right)
\left( t_{eq} \over 4.7 \times 10^4 {\rm yrs} \right)
\left( 1.35 \times 10^{10} {\rm yrs} \over t_{today} \right) \\
Y_0 & = & 104.4
\ea
which gives
\ba
\int l {dN \over dV dl} dl & \sim & {1 \over t^2}
\left( {\cal N}_1 \left( 
20.4 \sqrt{ Y \over Y_0 } -2 \right) 
+ {\cal N}_2 \left( 17.22 + \log \left( X \over X_0 \right) \right) \right) \\
{\cal N} & = & \frac {2 (2-\beta ) \Upsilon }
{\zeta  \left( 1 - {\left( \alpha_L \over \alpha_U \right)^{2 - \beta}} \right) } \\
\zeta & = & 4 - 3n - \beta 
\ea

Adopting numerical parameters approximately consistent with the
VS simulations
\ba
\alpha_U & =& 0.3 \\
\alpha_L & =& 10^{-4} \\
\beta & = & 1.6 \\
\Upsilon & =& 43.6
\ea
we find
\ba
{\cal N}_1 = 40.4 \\
{\cal N}_2 = 90.9
\ea
and
\ba
\int l {dN \over dV dl} dl & \sim & {2250 \over t^2} H(X,Y) \\
\Omega_{loops} & \sim & 8.5 \times 10^{-6} 
\left( G \mu \over 2 \times 10^{-10} \right)
H(X,Y) \\
H(X,Y) & \sim & \left( 0.63 + 0.37 \sqrt{ Y \over Y_0 } + 0.04  \log \left( X \over X_0 \right) \right)
\ea
where $X$ and $Y$ depend upon $\alpha_U/\Gamma_R G \mu$ and cosmological
timescales.

\subsection{$\Lambda$-CDM}

For a more realistic cosmology, consider a $\Lambda$-CDM model.
Let $\Omega_{r,0}=8.4 \times 10^{-5}$
$\Omega_{m,0}=0.3$ and $\Omega_{\Lambda,0}=1-\Omega_{r,0}-\Omega_{m,0}$.
The relationship between a and t is given by
\be
\int_0^a da {1 \over \sqrt{ \Omega_{r,0}a^{-2} + \Omega_{m,0}a^{-1} + \Omega_{\Lambda,0} } } = \tau
\ee
where $\tau(a) = H_0 t(a)$. We choose $a=1$ today ($t(1)=t_{today}$,
$H_0 = \tau(1)/t_{today}$) or $t(a) = t_{today} \tau(a)/\tau(1)$ and
$V(a) \propto a^3$.
Assuming that all parameterized quantities ($\beta$, $\alpha_U$,
$\alpha_L$, $\alpha_G$, $\gamma_s$) are constant,
we evaluate the loop density at $t_{today}$ numerically
\be
{dN \over dV dl} = {A \over l^\beta V(t_{today})} \int_{t_0^*}^{t_1^*} dt' 
t'^{\beta-5} V(t')
\ee
where
\be
t_1^* = \min{\left( t_{today},{l \over \alpha_L} \right) }
\ee
\be
t_0^* = \max{\left( 0,t_{today}-{l \over \alpha_G},{l \over \alpha_U} \right) } .
\ee
These results are included on the plots; they are the solid, smooth
lines. They agree fairly well with the simple, two-powerlaw
approximation.

\subsection{$\Lambda$-CDM: Numerical Results for varying $G \mu$, $\alpha_U$,
$\alpha_L$ and $\beta$}

We choose a fiducial set of parameters: $\beta = 1.6$, $\Gamma_R G \mu = 
10^{-8}$, $\alpha_U = 0.3$ and $\alpha_L=10^{-4}$.  We then
varied a single parameter over a wide range to explore how the loop
distribution function changed.

Effect of variation of $G \mu$ on the loop distribution is illustrated in
Figure \ref{fig:Lambda-mu2}. Decreasing the tension extends the
distribution to smaller size scales because gravitational
radiation is less rapid.
\FIGURE{
\epsfig{file=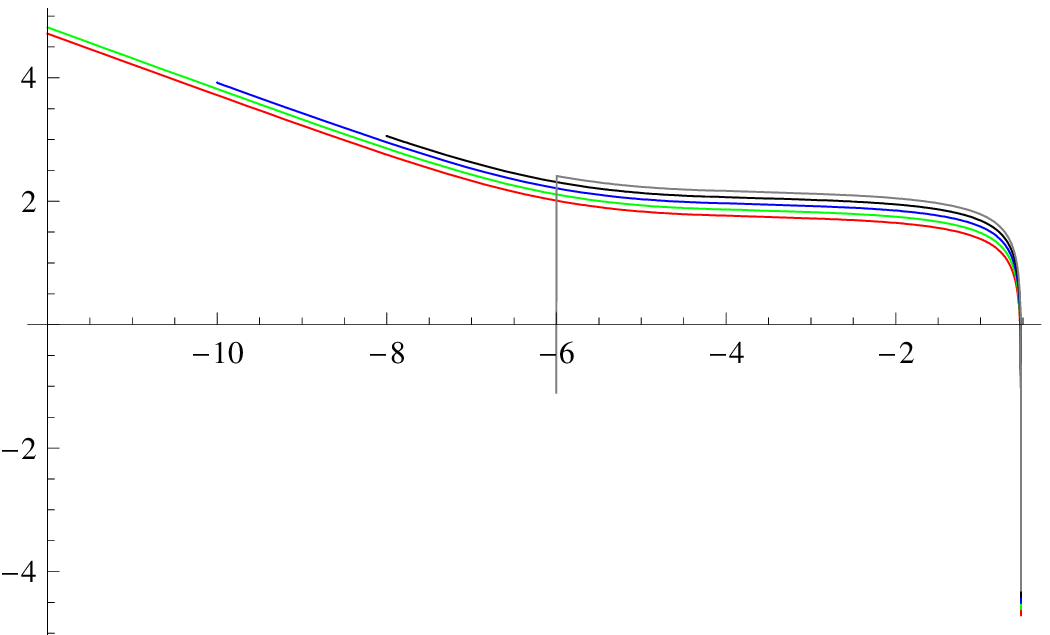}
\caption{Loop distribution for $\Gamma_R G \mu = 10^{-15}$ to $10^{-9}$
in steps of $10^2$ for standard case. Slight offsets to expose separate
curves. Lower $\Gamma_R G \mu$ extend to smaller loop size.}
\label{fig:Lambda-mu2}
}

Effect of variation of $\beta$ on the loop distribution is illustrated in
Figure \ref{fig:Lambda-beta2}. Smaller $\beta$ concentrates loops at the largest
scale relative to the horizon. These large loops live longer 
with respect to gravitational wave damping modestly increasing the
abundance of the smallest loops.
\FIGURE{
\epsfig{file=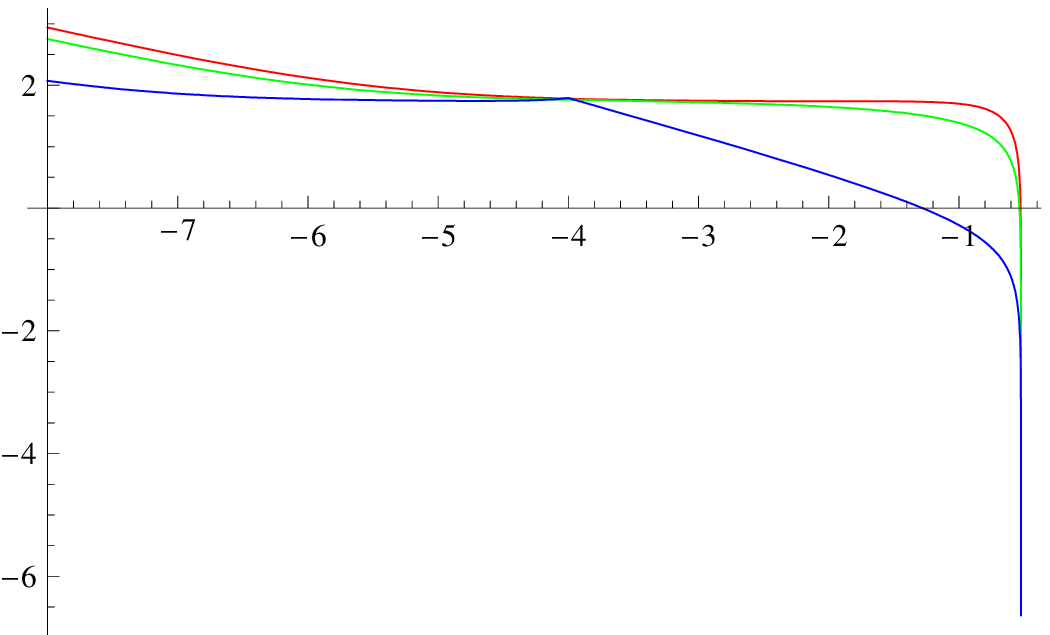}
\caption{Loop distribution for $\beta = 0.6$ (red), $1.6$ (green) and $2.6$
(blue).}
\label{fig:Lambda-beta2}
}

Effect of variation of $\alpha_U$ on the loop distribution is illustrated in
Figure \ref{fig:Lambda-alphamax2}. Smaller $\alpha_U$ limits the size of the
loops at all times and leads to more rapid decay.
\FIGURE{
\epsfig{file=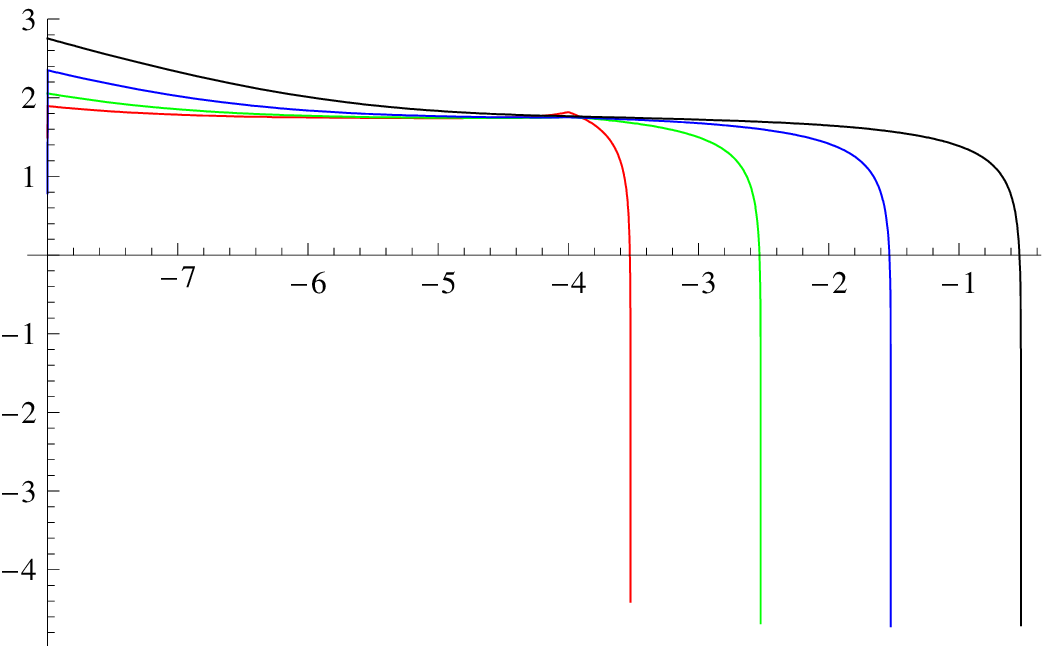}
\caption{Loop distribution for $\alpha_U = 3 \times 10^{-4}$ (red), 
$3 \times 10^{-3}$ (green),
$3 \times 10^{-2}$ (blue),
$0.3$ (black).}
\label{fig:Lambda-alphamax2}
}

Effect of variation of $\alpha_L$ on the loop distribution is illustrated in
Figure \ref{fig:Lambda-alphamin2}. Larger $\alpha_L$ increases the size of the
loops at all times and leads to less rapid decay.
\FIGURE{
\epsfig{file=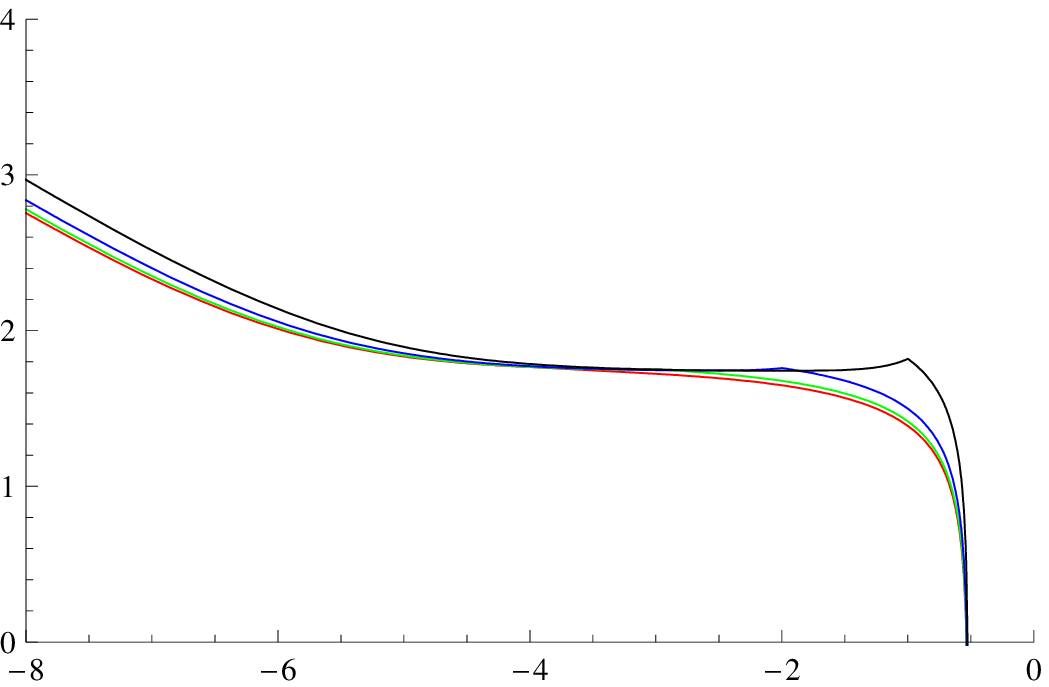}
\caption{Loop distribution for $\alpha_L = 10^{-4}$ (red), 
$10^{-3}$ (green),
$10^{-2}$ (blue),
$0.1$ (black).}
\label{fig:Lambda-alphamin2}
}

\end{document}